\documentclass[aps,twocolumn,showpacs,showkeys,amsmath,amsfonts,amssymb]{revtex4}
\usepackage{graphicx}
\usepackage{epsf}
\begin{document}

\title{Gapped solitons and periodic excitations in strongly coupled BEC}

\author{ Utpal Roy }
\email{utpal.roy@unicam.it}
\affiliation{Dipartimento di Fisica, Universit\`{a} di Camerino, I-62032 Camerino, Italy}
\author{B. Shah }
\email{balkrishna_29@yahoo.com}
\affiliation{Department of Physics, Faculty of Science, M. S. University of Baroda, Vadodara-390002, Gujarat, India}
 
\author{Kumar Abhinav}
\email{kumarabhinav@iiserkol.ac.in}
\author{P. K. Panigrahi}
\email{pprasanta@iiserkol.ac.in}

\affiliation{Indian Institute of Science Education and Research-Kolkata, Mohanpur, Nadia-741252, West Bengal, India}

\begin{abstract}

It is found that localized solitons in the strongly coupled cigar shaped Bose-Einstein condensate form two distinct classes. The one without a background is an asymptotically vanishing, localized soliton, having a wave-number, which has a lower bound in magnitude. Periodic soliton trains exist only in the presence of a background, where the localized soliton has a \textit{W}-type density profile. This soliton is well suited for trapping of neutral atoms and is found to be stable under Vakhitov-Kolokolov criterion, as well as numerical evolution. We identify an insulating phase of this system in the presence of an optical lattice. It is demonstrated that the ${\it W}$-type density profile can be precisely controlled through trap dynamics.

\end{abstract}

\maketitle

\section{Introduction}

Cigar shaped Bose-Einstein condensate (BEC) provides an ideal venue for realizing different types of solitonic excitations in one dimension. Dark \cite{Burger,Denschlag}, bright \cite{Khaykovich,Strecker,Khawaja,Cornish}, grey \cite{Shomroni} solitons and periodic soliton trains \cite{Strecker} have been experimentally observed. In the weak coupling regime, the Gross-Pitaevskii (GP) equation, describing the mean field BEC dynamics, reduces to the well studied non-linear Schr\"{o}dinger equation (NLSE), possessing soliton solutions. The fact that NLSE is an integrable system explains the observed soliton dynamics \cite{Zakharov1, salasnich1, mateo, Zakharov}. In one dimension, the possibility of using these localized excitations for producing atom laser and other applications \cite{ketterle, Pasquini, Billy}, requiring macroscopic coherence, has made cigar-shaped BEC an area of significant current interest \cite{leggett,anglin,stringari,pethick,CrExp}. The fact that scattering length can be controlled through Feshbach resonance \cite{Stenger}, has given access to both weak and strong coupling sectors, as well as attractive and repulsive domains. The regulation of the transverse trap frequency ${\omega}_{\perp}$ can also be used to control the coupling in the cigar-shaped BEC, as has been demonstrated experimentally in the generation of the Faraday modes \cite{Engels}.

The strongly coupled cigar shaped BEC, in the repulsive domain, is characterized by a quadratic non-linearity, following the Thomas-Fermi approximation \cite{jackson1,salasnich}. In contrast to the weak coupling domain, the solitons in the strong coupling sector are not well studied. The grey soliton dynamics has been numerically investigated, which yielded a structure similar to that of the weak coupling regime \cite{Jackson2}. Here, we study dark solitons, as well as soliton trains in this non-linear system and find significant differences, with the weak coupling sector.  We restrict ourselves to the repulsive domain, keeping in mind the fact that strong coupling BEC will be prone to instability in the attractive sector. 

It is observed that unlike the case of NLSE, the solitons exist in two distinct classes. Starting from a general ansatz solution with non-vanishing background, it is found that there can be two domains of localized solutions. The one with background is a W-type soliton, which can exist for all the values of the momentum of the envelope profile. The other solution is an asymptotically vanishing soliton, without a background, which requires a finite momentum to get excited. Both the solutions are shown to be stable under the Vakhitov-Kolokolov (VK) criterion \cite{VKC}. It is observed that dependence of the effective chemical potential on the profile width separates these two solution domains and also ensures their stability. The \textit{W}-type soliton is shown to be dynamically stable under the numerical Crank Nicholson finite difference method. Interestingly, periodic cnoidal waves can only exist with a background. Using a general Pad\'{e} type ansatz, we identify more general solutions. It is shown that the background must be real and non-vanishing for the W-type solitons to exist. Since W-type soliton is well-suited for trapping of atoms, we illustrate the procedure for coherent control of this structure, through scattering length and trap frequency.  

The paper is organized as follows. In Sec.II, we obtain exact localized and periodic solutions of the strongly coupled GP equation in one dimension, and compare them with the weak coupling sector. In Sec.III, a more general ansatz is employed to identify some of the non-linear excitations, unavailable through standard procedure. It is found that a unique localized solution can exist only with a real background. We also explore the structure of the solutions in the presence of an optical lattice and identify an insulating phase. Subsequently in Sec.IV, their coherent control is analytically demonstrated, where both the amplitude and width of the \textit{W}-type soliton can be exactly regulated. We conclude in the final section with a number of directions for future work.

\section{Exact solitons of the strongly coupled BEC}

The three dimensional GP equation describing the dynamics of BEC in
a cylindrical harmonic trap, $V=\frac{1}{2}M\omega^2_\bot(x^2+y^2)$,
is given by

\begin{equation}
\label{GP} i\hbar \frac{\partial \Psi({\bf{r}},t)}{\partial
t}=\left[-\frac{\hbar^2}{2M}\nabla^{2}+U_{0}|\Psi({\bf{r}},t)|^2+V\right]\Psi({\bf{r}},t),
\end{equation}

where $U_0=4\pi\hbar^2a/M$, is the effective
two body interaction, $a$ being the scattering length. In the
quasi-one dimensional limit, the condensate wave-function can be factorized:

\begin{equation}
\Psi({\bf{r}},t)=f(z,t)G(x,y,\sigma),
\end{equation}
where $\sigma(z)$ is the local particle density. $G(x,y,\sigma)$ is
the normalized equilibrium wave function for the transverse motion:

\begin{equation}
\sigma(z)=\int{dx dy |\Psi(x,y,z)|^2}=|f(z,t)|^2.
\end{equation}

 In the repulsive, strong coupling limit ($n_0 U_0>>\hbar
\omega_{\perp}$, where $\omega_{\perp}$ is the confining trap frequency), one uses the Thomas-Fermi approximation for the
transverse profile, leading to the condensate equation
\cite{jackson1},

\begin{widetext}
\begin{eqnarray}
i\hbar\frac{\partial}{\partial
t}f(z,t)=\left[-\frac{{\hbar}^2}{2M}\frac{\partial^2}{\partial z^2}+ 2\hbar \omega_\perp a^{1/2}\left(|f(z,t)|-{\sigma_0}^{1/2}\right)\right]f(z,t). \label{NLSE}
\end{eqnarray}
\end{widetext}

Here, $\sigma_0$ is the equilibrium density of the atoms far away
from the axis. The non-linear excitations of this system can be probed through an ansatz solution,

\begin{equation}
f(z,t)=e^{i(kz-\omega t)}\rho(\xi),
\end{equation}

with a fast moving component and slowly varying envelope profile
$\rho(\xi)$. Here, $\xi=\alpha(z-vt)$ and $v=\frac{\hbar}{M}k$, with
$\rho(\xi)$ satisfying,

\begin{equation}
\alpha^2\rho''+g\rho^2+\epsilon\rho=0 \label{real},
\end{equation}

where $g=-4M\omega_\perp a^{1/2}/\hbar$ and
$\epsilon=2M\omega/\hbar+4M\omega_{\bot}(\sigma_0
a)^{1/2}/\hbar-k^2$. 

It is straightforward to check that the following ansatz solution,
\begin{equation}
\rho(\xi)=A+Bcn^{2}(\xi ,m),
\end{equation}

solves Eq.6, where $cn(\xi ,m)$ is the cnoidal function, with $m$ being  the modulus parameter $(0\leq m \leq 1)$ \cite{Hancock},  and $A,B$ are constant parameters to be determined. We note that $cn(\xi , 0)=cos(\xi)$ and $cn(\xi, 1)=sech(\xi)$. $A$ serves here as the background.  On substitution, we arrive at the \textit{consistency conditions},

\begin{eqnarray}
-2B{\alpha}^{2}(m-1)A+gA^2+\epsilon A &=& 0,\nonumber \\
-4{\alpha}^{2}(1-2m)+2gA+\epsilon &=& 0,\nonumber \\
-6m{\alpha}^{2}+gB &=& 0,
\end{eqnarray}

leading to,

\begin{eqnarray}
A &=& \frac{1}{2g}\left[4{\alpha}^{2}(1-2m)-\epsilon\right], \nonumber \\
B &=& \frac{6}{g}{\alpha}^2 m, 
\end{eqnarray}

along with a relation between the width $\alpha$, and the effective chemical potential $\epsilon$,

\begin{equation}
{\epsilon}^2 = 16{\alpha}^{4}\left(m^2 - m +1\right).
\end{equation}

The sign of $\epsilon$ is not fixed, as both the roots of the above equation are allowed. The general periodic solution can then be written as,

\begin{eqnarray*}
\rho(\xi)=\frac{2{\alpha}^{2}}{g}\left[(1-2m)\mp\sqrt{m^2-m+1}+3m{cn}^{2}(\xi ,m)\right],
\end{eqnarray*}

representing a soliton train.  
Consideration of the special case $m=1$ yields localized solutions. It corresponds to two specific values, $\epsilon= \pm 4{\alpha}^2$. The \textit{positive} root requires presence of the background $A$, and the envelope takes the form, 

\begin{equation}
\rho(\xi)=-\frac{\epsilon}{g}\left[1-\frac{3}{2}sech^2(\xi)\right],
\end{equation} 

representing a localized \textit{W}-type soliton.

The Vakhitov-Kolokolov criterion points out that the integral $N(\epsilon)=\int |\rho(\xi)|^{2}d\xi$, when varied with respect to the \textit{effective} chemical potential $\epsilon$, indicates the stability of the solution $\rho(\xi)$. In the present case, 

\begin{eqnarray*}
\frac{dN(\epsilon)}{d\epsilon}=-\frac{6\epsilon}{g^2},
\end{eqnarray*}

requiring that $\epsilon >0$ for the stability of the solution, which is consistent with $\epsilon =4{\alpha}^2$.\\
 
For this case we find,

\begin{equation}
k^2\geq 2\frac{M\omega}{\hbar}-|\epsilon|,
\end{equation} 

setting a lower limit $\frac{\hbar |\epsilon|}{2M}$ for $\omega$, if it is positive, in order to generate the \textit{W}-type soliton. Such bound is not there if the driving frequency is negative.\\

For the \textit{negative} root of Eq.10, with $m=1$, the background vanishes, and we get,

\begin{equation}
\rho(\xi)=-\frac{3\epsilon}{2g}sech^{2}(\xi),
\end{equation}

which, under the Vakhitov-Kolokolov criterion, yields,

\begin{eqnarray*}
\frac{dN(\epsilon)}{d\epsilon}=6\frac{\epsilon}{g^2},
\end{eqnarray*}

and hence is stable as $\epsilon <0$. In this case, $k^2\geq |\epsilon|+2\frac{M\omega}{\hbar}$, indicating that a finite wave number is needed to excite the solution for the positive frequency case. Such condition does not arise if the driving frequency is negative. This type of velocity restricted solitons have been identified in higher order non-linear Schr\"{o}dinger equation relevant for optical fiber pulses in the femtosecond domain \cite{Vivek}.

\begin{figure}[htpb]
\includegraphics[width=2.2 in]{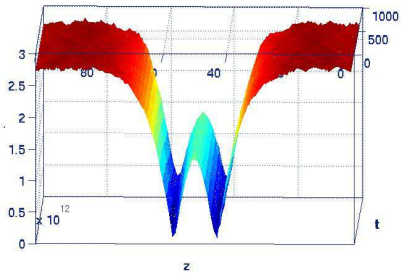}
\caption{(Color Online) Direct numerical evolution of ${\it W}$ type
soliton with $dz=0.0003$, $dt=0.0001$ for $1000$ iterations.}
\label{crank}
\end{figure}

We have numerically evolved the \textit{W}-type solution, using the
Crank– Nicholson finite difference method, which is unconditionally
stable. In this analysis, the initial profile has been taken as
$\psi(z, t = 0) = \psi(z, t = 0) + \tilde\epsilon$, where $\tilde\epsilon$ is a
function, which assumes a random value at each point. The analysis
was carried out with $dz=0.0003$, $dt=0.0001$ for $1000$ cycles.
Fig.~\ref{crank} shows that, the \textit{W}-type soliton remains unchanged with its form showing minor  perturbation, where $\tilde\epsilon$ is taken to be $10$ percent of the peak value of $\psi$. The minima positions and the
width remained unaltered. We have also checked that the evolution was unitary conserving the
number of particles, upto second order in $dt$.\\

\section{Complex Envelope Bloch solitons}

A general ansatz for identifying complex envelope Bloch type solitons can be written as,

\begin{equation}
\rho(\xi)=(A+iB)+(C+iD)cn^{2}(\xi ,m),
\end{equation}

where $A$,$B$,$C$,$D$ are real. The corresponding consistency conditions lead to,

\begin{widetext}
\begin{eqnarray}
g(A+iB)\sqrt{A^2+B^2}-\epsilon(A+iB)+2{\alpha}^2(C+iD)(m-1)&=&0,\nonumber \\ \nonumber \\
\left[g^2(A^2+B^2)-{\epsilon}^2\right](A+iB)(C+iD)+g^2(AC+BD)(A+iB)^2-8{\alpha}^4(C+iD)^2(m-1)(1-2m)\nonumber \\ 
-2\epsilon{\alpha}^2\left[(C+iD)^2(m-1)+2(A+iB)(1-2m)(C+iD)\right]&=&0,\nonumber \\ \nonumber \\
g^2\left[(C^2+D^2)(A+iB)^2+(A^2+B^2)(C+iD)^2+4(AC+BD)(A+iB)(C+iD)\right]-{\epsilon}^2(C+iD)^2\nonumber \\
-4\epsilon{\alpha}^2(C+iD)\left[3(A+iB)m2(C+iD)(1-2m)+\right]-8{\alpha}^4(C+iD)^2\left[2(1-2m)^2+3m(m-1)\right]&=&0,\nonumber \\ \nonumber \\
g^2\left[(AC+BD)+(C-iD)(A+iB)\right]-6\epsilon{\alpha}^2m-24{\alpha}^4m(1-2m)&=&0,\nonumber \\ \nonumber \\
g^2(C^2+D^2)-36{\alpha}^4m^2&=&0.
\end{eqnarray}
\end{widetext}

Solutions of these consistency conditions can be found through a tedious, but straightforward calculation;

\begin{widetext}
\begin{eqnarray}
A&=&\sqrt{p -B^2}, C=\sqrt{q^2-\left(\frac{r}{p}\right)^2B^2}, D=\frac{r}{p}B,\nonumber \\
B^2&=&\left[p(g\sqrt{p}+\epsilon)^2-4r{\alpha}^4(m-1)^2\right]/\left[(g\sqrt{p}+\epsilon)^2-4{\alpha}^4(m-1)^2(r/p)^2\right];
\end{eqnarray}

where,

\begin{eqnarray}
p&=&\frac{1}{3g^2}\left[{\epsilon}^2+6(1-2m)\epsilon{\alpha}^2+4{\alpha}^4(11m^2-11m+2)\right],\nonumber \\
q&=&\frac{36{\alpha}^4{m}^2}{g^2}, r=\frac{3{\alpha}^2m}{g^2}\left[4{\alpha}^2(1-2m)+\epsilon\right];
\end{eqnarray}

\end{widetext}

and, ${\epsilon}^2=16{\alpha}^4$. In case of localized solitons, i.e., for $m=1$, we arrive at the interesting condition that $A=0=C$, leading to the previously found \textit{W}-type soliton. However, periodic cnoidal wave solutions can exist with complex background parameters.

We now explore more general solution space, through a Pad\'{e} type ansatz \cite{raju}, 

\begin{equation}
\rho(\xi)=\frac{A+B\;f(\xi)}{1 + C\;f(\xi)}.
\label{generalFT}
\end{equation}

$A,\;B,$ and $C$ are real parameters to be determined from the consistency conditions, with $AC-B\neq 0$. It can be seen that these conditions
are four in number, indicating the constrained nature of the
general solution. The following consistency conditions,
with $\alpha^2\equiv\beta$ can be deduced:

\begin{widetext}
\begin{eqnarray}
A^2g+2BC\left(m-1\right)\beta +A\left(-2\,C^2\left(m-1\right)
\beta+\epsilon\right)&=&0,\nonumber\\
A^2Cg+B\left(\left(2m-1\right)\beta+\epsilon\right)
+A\left(2Bg+C\left(\beta-2m\beta+2\epsilon\right)\right)&=&0,\nonumber\\
B^2g+AC^2\left(-\beta+2m\beta+\epsilon\right)+
BC\left(2Ag+\beta-2m\beta+2\epsilon\right)&=&0,\nonumber\\
B^2Cg-2Bm\beta +2ACm\beta +BC^2\epsilon&=&0 \label{linearconsistency}.
\end{eqnarray}
\end{widetext}

We concentrate on the localized solutions, because
of its physical interest:

\begin{equation}
\rho(\xi)=-\frac{\epsilon}{g}\left(\frac{1- 2sech(2\xi)}{1+
sech(2\xi)}\right).\label{nss1}
\end{equation}

Like the previous case, the width and the amplitude of the
solution are coupled. It is easy to see that the profile can be rewritten as,

\begin{equation}
\rho(\xi)=-\frac{\epsilon}{g}\left(1- \frac{3}{2}sech^2
(\xi)\right),\label{LFTlin}
\end{equation}

 which is identical to the \textit{W}-type soliton.
 
  The above solution can be identified as the unique separatrix in the phase space of the solutions of Eq.(6), separating the periodic solutions with closed orbits from the unbounded ones, represented by open orbits \cite{VivekR}. \\ 

The presence of periodic solutions motivates us to explore the nature of the solution in the presence of an optical lattice,
$V(x) = V_{0} \cos^{2}x$, which necessitates the solutions to possess sinusoidal character. For the purpose of comparision \cite{Priyam}, the 1-D GP equation is normalized in the form,

\begin{equation}
i \partial_{t}\psi(z,t) = ( -\frac{1}{2} \partial_{zz} + g
|\psi(z,t)| + V(x) - \mu )\psi(z,t), \label{GP_lattice}
\end{equation}

where, $g$ is the normalized two-body interaction. The solution is found to be of the form,
$\psi(z,t) = (a + b \cos^{2}z) e^{- i \omega t}$. The parameter values for this insulating phase is given by, $a = V_{0}/{2 g} \textrm{\hskip 0.5cm and\hskip 0.5cm} b =
-V_{0}/g$, with $\omega = 1 + 1/2\; V_{0} - \mu$. In contrast to the weak coupling case
\cite{carr}, where, analytic solutions have been obtained for both superfluid and
insulating phases, for the present case, we have been able to identify only an insulating
phase, analytically. It is worth observing that in this case, the competition between the lattice
potential and the nonlinearity yields sinusoidal solutions,
whereas, for the localized soliton solutions, the
nonlinearity and the dispersion are responsible for the existence
of the solutions. In the repulsive domain, the above solutions
exist both for positive and negative values of $V_0$.
The corresponding energy (in dimensionless
unit) is found to be $ E = 4d \pi
V^{2}_{0} + d \pi V^{3}_{0} - 2d \pi \mu V^{2}_{0}$, where $d=1/{8
g^{2}}$. We note that the contribution from the interaction
term does not explicitly contribute to the energy, although the coupling
parameter appears in $E$ through the solutions. The average atom
number density, $\nu = \frac{V^{2}_{0}}{8 g^{2}}$, is a
constant and can be controlled by tuning the lattice potential and
the scattering length.

\section{Coherent control of the solitons}

 We now investigate the effect of time dependent nonlinearity and gain or loss on the \textit{W}-type soliton profile. The two points of this soliton, where the order parameter vanishes, may be useful for trapping of neutral atoms. The barrier height and the locations of the minima can be controlled, by changing the frequency and the scattering length, which can be manipulated through Feshbach resonance \cite{donley,papp}. Keeping this in mind, in the folloeing, we study the control of this localized excitation in the following section. For the sake of comparison with the weak coupling case \cite{atre, Ramesh, Sree}, the GP equation is cast in the form

\begin{equation}
i \partial_t \psi=-\frac{1}{2}\partial^2_{zz}
\psi+\gamma(t)|\psi|\psi+\frac{1}{2}M(t)z^2\psi+
\frac{i\kappa(t)}{2}\psi. \label{NLSE1}
\end{equation}

Here $\psi$, $t$ and $z$ have been scaled, respectively by $a_B
^{1/2}$, $\omega_\perp$ and $1/a_\perp$, making them dimensionless.
$\gamma=2(a(t)/a_B)^{1/2}$ is time dependent non-linearity
coefficient, controllable through Feshbach resonance;
$M(t)=\omega_{0}^{2}(t)/\omega_{\perp}^{2}$ is related to the axial
trap frequency, which can be made time dependent.
$\kappa(t)=\eta(t)/\hbar \omega_{\perp}$ is time dependent
loss/gain. The oscillator length in the transverse direction is
defined as $a_{\perp}=(\hbar/M\omega_{\perp})^{1/2}$ and $a_B$ is
the Bohr radius. We consider the following ansatz solution for:

\begin{equation}
\label{ansatz}
\psi(z,t)=B(t)F
(\xi)e^{\left[i\Phi(z,t)+\frac{1}{2}G(t)\right]},
\end{equation}

where $T=A(t)\{z-l(t)\}$, $l(t)=\int ^{t}_{0}v(t^{'})dt^{'}$ and
$G(t)=\int ^{t}_{0}\kappa(t^{'})dt^{'}$. The phase has been taken in
the form $\Phi(z,t)=a(t)+b(t)z-\frac{1}{2}c(t)z^2$, where $a(t)$ is
a $z$ independent phase term:
$a(t)=a_{0}+\frac{\lambda-1}{2}\int_{0}^{t}A^{2}(t')dt'$. The
solutions are necessarily chirped in time and space, with time
varying amplitude and width. $b(t)$ is a time dependent momentum and
$c(t)$ balances the oscillator, leading to the Riccati equation:

\begin{equation}
c_{t}-c^{2}(t)=M(t).
\end{equation}
The above can be cast as the familiar Schr\"{o}dinger eigen value
equation:
\begin{equation}
-\phi^{''}(t)-M(t)\phi(t)=0,
\end{equation}

via a change of variable: $c(t)=-\frac{\partial ln\phi(t)}{\partial
t}$. The constant part of $M(t)$ acts as the eigen value. A number
of variations in the oscillator frequency can be analytically
incorporated from solvable quantum mechanical problems. The
oscillator can also be made expulsive. The location of the
condensate profile satisfies, $dl(t)/dt-c(t) l(t)=b(t)$. Other
parameters are obtained, using the following consistency conditions:
$B(t)=B_0 \;exp(1/2 \int_{0}^{t} c(t^{'}) dt^{'})$,
$\gamma(t)=-\frac{1}{2}\gamma_0\;A(t)B(t)exp(-G(t)/2)$, and $
A(t)=B^2 (t)=b(t)$. The real part of the GP equation yields,

\begin{equation}
F^{''}(T)+gF^2 (T)+\epsilon F(T)=0 \label{real1},
\end{equation}

which takes the form of Eq.~\ref{real}, with
$g=-2\frac{\gamma(t)}{A(t)B(t)}e^{G(t)/2}$, $\epsilon=-\lambda$ and
$T=A(t)[z-l(t)]$. One can find various singular and non-singular
solutions for $F(T)$, using the previous procedure, provided $g$ is
constant. We now illustrate two specific cases of
interest.

I. First, we consider the condition $M(t)={M_0}^2$, with a cnstant oscillator frequency. This yields a periodic solution, with period $\pi/2M_0$;

\begin{eqnarray}
\psi(z,t)=-\frac{\epsilon}{g}&\sqrt{A_0 sec(M_0 t)}&[1-
\frac{3}{2}sech^2 (T/2)]\nonumber\\
&\times&e^{i\Phi(z,t)+\frac{1}{2}G(t)}
\end{eqnarray}

\begin{figure}
\centering
\includegraphics[width=3. in]{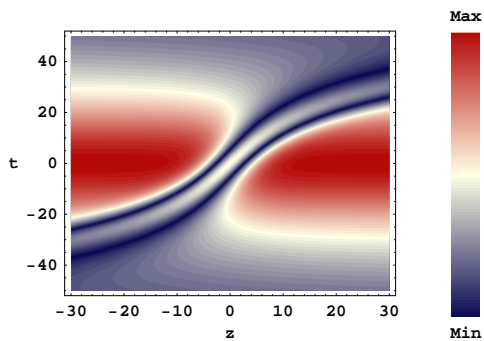}
\caption{(Color Online) Temporal behavior of ${\it W}$ type soliton,
without gain/loss and with $M_0=0.5,\;A_0=.01,\;\lambda=-1$ and
$\gamma_0=10$.} \label{dist12}
\end{figure}

The soliton can be compressed and accelerated through the time dependence of coupling. The presence of $sec(M_0t)$ in the amplitude and width of the soliton profile, leads to its compression and amplification, as time increases.

II. We now turn our attention to the solution in an expulsive
potential ($M(t)=-M_0 ^2$), where the ${\it W}$ type density profile
is given by,

\begin{eqnarray}
\psi(z,t)=-\frac{\epsilon}{g} &\sqrt{A_0 sech(M_0 t)}& [1-
\frac{3}{2}sech^2 (T/2)]\nonumber\\ &\times & e^{i\Phi(z,t)}.
\end{eqnarray}

This profile has a transient character and the behavior in time for the expulsive potential
is very different from the regular case.

The fact that these minima
locations can be controlled, including the barrier height, makes
these solutions potentially attractive, for trapping of neutral atoms and their manipulation.
The temporal evolution of the density profile is shown in Fig.2. One observes gradual reduction in amplitude as time progresses.

\section{Conclusion}
In conclusion, the cigar shaped BEC in the strong coupling sector leads to two different types of stable localized solitons, absent in the weak coupling regime. Localized solitons, with no background, require a finite momentum to exist for positive driving frequency. No such restrictions are there for the soliton trains which can propagate only on a background. The solutions are found to be stable under VK criterion. It is shown that the localized soliton can be effectively controlled through, scattering length and trap frequencies, which makes it useful for trapping of atoms. It is interesting to note that the asymptotically vanishing localized solution is also velocity restricted. This type of solutions exist in higher order non-linear equations, relevant for femto-second pulses in optical fiber \cite{Vivek}. We hope that the velocity restricted solitons, so far unobserved in the optical fiber system, may find experimental verification in cold atoms. It will be of interest to find exact solutions for the non-polynomial mean field equation relevant for cigar shaped BEC \cite{salasnich} and compare the transition between strong and weak coupling sectors. It will be of deep interest also to extend the present procedure to the mean field equations governing a Boson-Fermion mixture of atomic gases \cite{Beitia}.

\end{document}